\documentclass[12pt,epsfig]{article}
\usepackage{amsfonts,amssymb,amsmath,amscd,dsfont,wrapfig,graphicx}
\usepackage[                                         
  textwidth=16cm,
  outer=2cm,
  textheight=45\baselineskip,
  headheight=\baselineskip,
  includehead=true,
  heightrounded,
]{geometry}
\DeclareFontFamily{U}{rsfs}{}         
\DeclareFontShape{U}{rsfs}{m}{n}{<5> rsfs5 <6><7> rsfs7          %
  <8><9><10><10.95><12><14.4><17.28><20.74><24.88> rsfs10}{}     %
\DeclareMathAlphabet{\mathfs}{U}{rsfs}{m}{n}                     %
\newcommand{\mfs}[1]{\mathfs {#1}}                               %
\DeclareFontFamily{OT1}{pzc}{}
\DeclareFontShape{OT1}{pzc}{m}{it}{<-> s * [0.900] pzcmi7t}{}
\DeclareMathAlphabet{\mathscr}{OT1}{pzc}{m}{it}

\input{epsf}
\def\beq{\begin{eqnarray}}
\def\eeq{\end{eqnarray}}
\def\bea{\begin{align}}
\def\eea{\end{align}}

\def\nn{\nonumber\\}
\def\nd{\noindent}

\def\del{\partial}
\def\grad{\nabla}

\def\vev#1{\langle#1\rangle}

\def\bx{{\bf x}}

\def\bk{{\bf k}}

\title{HAWKING RADIATION -- REVISITED}
\author{Amit Ghosh\footnote{amit.ghosh@saha.ac.in}\\ 
\small Saha Institute of Nuclear physics \& Homi Bhabha National Institute\\
\small 1/AF, Bidhan Nagar, Kolkata 700064, India\\}

\begin{document}
\maketitle

\begin{abstract}
In this paper we revisited Hawking radiation in the light of the original calculations of Hawking and Wald and found that some additional insights can be gained. We review a ``derivation" of the field theory Hilbert space from the space of solutions, followed by the calculation of Bogoliubov coefficients in a collapsed spacetime from first principles. We show that these calculations can be generalized to the case of local Killing horizons and also to asymptotically non-flat spaces.   
\end{abstract}

\newpage\normalsize\pagenumbering{arabic}

\section{Introduction}

Even after many years of its discovery \cite{HAWKING}, Hawking radiation did not loose its significance or relevance, see \cite{Visser}-\cite{Singleton}. The reason for its importance is manifold. Not only its existence has been observed in systems that are far from being black holes \cite{Visser}, but it raised several critical issues in black holes also \cite{Wald2}-\cite{Brout}. The original work \cite{HAWKING} was soon extended to a full scale quantum field theory calculation in \cite{Wald1}. One of the difficulty was that the set up used in the original works was too global, it made use of event horizons and global techniques. On the other hand, observational physics usually depends on local properties. Hence it is important to be able to carry out the analysis, expecting that the original results will still hold, using local techniques and local structures of space-time. Usually, a black hole is best characterised by its horizon and not by its asymptotic structures. So it is reasonable to attempt a derivation of Hawking radiation using germs of horizon only, irrespective of its behavior at asymptotic infinities. It inspired a huge set of fresh investigations, see \cite{Parikh}-\cite{Wilczek}. This field is still active. However, in this work we have tried to follow the original derivation as closely as possible. We find that there exists a small window through which the original techniques can be extended to local neighbourhoods of the horizon, avoiding its asymptotic properties. We closely followed the quantum field theory techniques developed by Geroch \cite{Geroch} and used in \cite{Wald1}. We elaborated some of the steps to make the framework more lucid so that a wider set of people can use it. Our discussion (towards the end of section 2) on how to generalize these techniques to a full scale quantum field theory in curved space-time is, however, too inadequate and limited. Readers who want to get a thorough and up-to-date knowledge of the subject should see \cite{Wald3}. We elaborated some calculations to show how the same formulas can be made to work in local context. These calculations like the explicit maps between the solutions and the Hilbert spaces are presented in section 3 and 4. We gave a brief conclusion at the end and left some open question.    

The Hilbert space of a free quantum field is an infinite direct sum of many-particle Hilbert spaces ${\bf F}={\bf H}_0\oplus{\bf H}_1\oplus{\bf H}_2\oplus...$ where ${\bf H}_n$, $n=1,2,...$, is a $n$-particle Hilbert space and ${\bf H}_0$ is the vacuum, a 1-dimensional Hilbert space isomorphic to $\bf C$ that contains no particles. If $\bf H,H'$ are two Hilbert spaces with $\vev{\psi|\phi},\vev{\psi'|\phi'}'$ as their respective inner products then their direct sum $\bf H\oplus H'$ is a Hilbert space with the inner product
\beq \vev{\psi,\psi'|\phi,\phi'}=\vev{\psi|\phi}+\vev{\psi'|\phi'}'.\label{directsum}\eeq
It is straightforward to verify that the inner product $\vev{\psi,\psi'|\phi,\phi'}$ is right linear, hermitian and positive. This roughly ensures that $\bf F$ is a Hilbert space (the infinite sum of complex numbers on the right of (\ref{directsum}) must suitably converge).

If ${\bf H}_n$, $n=2,3,...$, is a $n$-fold tensor product of ${\bf H}_1=\bf H$, namely ${\bf H}_n=\otimes_n\bf H$, then $\bf F$ is called the Fock space. If $\bf H,H'$ are two Hilbert spaces then their tensor product $\bf H\otimes H'$ is a Hilbert space with the inner product
\beq \vev{\psi\otimes\psi'|\phi\otimes\phi'}=\vev{\psi|\phi}\vev{\psi'|\phi'}'.\eeq
If the 1-particle Hilbert space $\bf H$ is separable having an orthonormal basis $e_i$ then a typical Fock space element is
\beq \Psi=(\Psi_0,\Psi_1,\Psi_2,\Psi_3,...)
=(\psi_0,\;\psi_ie_i,\;\frac 1{2!}\psi_{ij}e_i\!\otimes\!e_j,\;\frac 1{3!}\psi_{ijk}e_i\!\otimes\!e_j\!\otimes\!e_k,...)\label{fockstate}\eeq
where $\psi_0,\psi_i,\psi_{ij},...$ are arbitrary complex numbers that are totally symmetric for the bosonic states and totally antisymmetric for the fermionic states.
\vspace{0.5cm}

In $\bf F$, the 1-particle annihilation and creation operators, denoted by $a(\psi)$ and $a^\dagger(\psi)$ respectively for all 1-particle states $\psi\in\bf H$, are defined as follows: 

{\tt (1)} Both $a(\psi),a^\dagger(\psi)$ are linear operators in $\bf F$.

{\tt (2)} $a(\psi)$ is anti-linear in $\psi$ and $a^\dagger(\psi)$ is linear in $\psi$. 

{\tt (3)} Denote $a(e_i)=a_i$ and $a^\dagger(e_i)=a^\dagger_i$. 

{\tt (4a)} Bosonic case: Both $a_i,a^\dagger_i$ are derivations on tensors. For $\Psi$ in (\ref{fockstate}),
\begin{align} &a_i\Psi=(\psi_i,\;\psi_{ik}e_k,\;\frac 1{2!}\psi_{ikl}e_k\!\otimes\!e_l,\;...)\\
&a^\dagger_i\Psi=(0,\psi_0e_i,\;\psi_k e_{(i}\!\otimes\!e_{k)},\;\frac 1{2!}\psi_{kl}e_{(i}\!\otimes\!e_k\!\otimes\!e_{l)},\;...)\end{align}
where $e_{(i}\!\otimes\!e_{j)}=\frac 12(e_i\!\otimes\!e_j+e_j\!\otimes\!e_i)$. Straightforward calculations show that 
\beq a_ia_j=a_ja_i,\quad a^\dagger_ia^\dagger_j=a^\dagger_ja^\dagger_i,\quad a_ia^\dagger_j-a^\dagger_ja_i=\delta_{ij}.\eeq
Furthermore, $\vev{\Phi|a_i\Psi}=\vev{a^\dagger_i\Phi|\Psi}$, namely one is the formal adjoint of the other.

{\tt (4b)} In the fermionic case, both $a_i,a^\dagger_i$ are anti-derivations on tensors,
\begin{align} &a_i\Psi=(\psi_i,\;\psi_{ik}e_k,\;\frac 1{2!}\psi_{ikl}e_k\!\otimes\!e_l,\;...)\\
&a^\dagger_i\Psi=(0,\psi_0e_i,\;\psi_k e_{[i}\!\otimes\!e_{k]},\;\frac 1{2!}\psi_{kl}e_{[i}\!\otimes\!e_k\!\otimes\!e_{l]},\;...)\end{align}
where $e_{[i}\!\otimes\!e_{j]}=\frac 12(e_i\!\otimes\!e_j-e_j\!\otimes\!e_i)=e_i\wedge e_j$. Straightforward calculations show 
\beq a_ia_j=-a_ja_i,\quad a^\dagger_ia^\dagger_j=-a^\dagger_ja^\dagger_i,\quad a_ia^\dagger_j+a^\dagger_ja_i=\delta_{ij}.\eeq
Furthermore, $\vev{\Phi|a_i\Psi}=\vev{a^\dagger_i\Phi|\Psi}$, namely one is the formal adjoint of the other.

$N=\sum_ia^\dagger_ia_i$ is called the number operator as it measures the number of particles in each Hilbert space
\beq N\Psi=N(\Psi_0,\Psi_1,\Psi_2,\Psi_3,...)=(0\Psi_0,1\Psi_1,2\Psi_2,3\Psi_3,...)\eeq
In either case, bosonic/fermionic, $[N,a_i]=-a_i$ and $[N,a_i^\dagger]=a_i^\dagger$. In the Fermionic case, $N_i^2=N_i$ where $N_i=a^\dagger_ia_i$, namely each $N_i$ is a bounded operator. 

The basis $e_i$ is not unique. A unitary map $e_i\mapsto e'_i=Ue_i$ maps an orthonormal basis to another. It induces a unitary map $\tilde U$ in $\bf F$ as follows:
\beq \tilde U\Psi=\Psi'=(\psi,\;\psi_iUe_i,\;\frac 12\psi_{ij}Ue_i\!\otimes\!Ue_j,\;...)\eeq
such that $\vev{\tilde U\Psi|\tilde U\Phi}=\vev{\Psi|\Phi}$. The new annihilation operator $a'_i$ is to be compatible with the unitary map in the sense that $a'_i\Psi'=\tilde Ua_i\Psi$, which implies $a'_i=\tilde Ua_i\tilde U^\dagger$. 

\section{Scalar field}

A massive free scalar field is a hermitian operator $\phi$ on the Fock space and a generalized function that vanishes on test functions $\phi(g)=0$ of the form $g=(\grad^2-m^2)f$ where $f$ is an arbitrary test function. It implies at regular points $\phi(x)$ is a solution of the Klein-Gordon equation $(\grad^2-m^2)\phi(x)=0$. 

All solutions of the wave equation form a complex vector space $\bf S$. If $f_1,f_2$ are two complex solutions then a scalar product in $\bf S$ is defined by
\beq \vev{f_2|f_1}_{\rm KG}=i\int_\Sigma\overline{f_2}*df_1-f_1*d\overline{f_2}\label{solinnerp}\eeq
where $\Sigma$ is a smooth spacelike hypersurface. The scalar product $\vev{f_2|f_1}_{\rm KG}$ is independent of the choice $\Sigma$ because the difference of the scalar product for any two such choices $\Sigma_1,\Sigma_2$
\beq \int_{\Sigma_1-\Sigma_2}\overline{f_2}*df_1-f_1*d\overline{f_2}=\int_Vd(\overline{f_2}*df_1-f_1*d\overline{f_2})=0\eeq
where $V$ is a portion of the spacetime having $\Sigma_1-\Sigma_2$ as boundaries. The integral vanishes because of the KG equation $d*df\propto f$ and the identity $df\wedge*dg=dg\wedge*df$ for any two smooth functions $f,g$. The scalar product (\ref{solinnerp}) is linear in the right and anti-linear in the left. It is hermitian, $\overline{\vev{f_2|f_1}}_{\rm KG}=\vev{f_1|f_2}_{\rm KG}$, but not positive definite.

\nd Remarks: The independence of the choice of hypersurface $\Sigma$ fails to hold in presence of a potential $V(\phi)$ higher than quadratic. So (\ref{solinnerp}) can not be easily extended to the case of interacting fields. Instead, we can continue to use this scalar product for variations $\delta\phi$, that is in local tangent spaces of the space of solutions. 

In Minkowski space the mass hyperboloid $k^\mu k_\mu+m^2=0$ has two components, the positive mass-shell $\bf M_+$ on which $k^0=\omega=(\bk^2+m^2)^{1/2}$ and negative mass-shell $\bf M_-$ on which $k^0=-\omega$. In canonical coordinates, $k^\mu/m$ where $k^\mu=(k^0,\bk)$ is the unit normal to the hyperboloid which is timelike and future-directed on $\bf M_+$ and past-directed on $\bf M_-$. For obvious reasons, $\bf M_+$ is considered physical.

Consider the Klein-Gordon equation in Minkowski spacetime, $(\del^2-m^2)\phi(x)=0$. On $\bf M_\pm$, the induced metric is $g_{ij}=\delta_{ij}-k_ik_j/\omega^2$; the Riemann volume element is $(m/\omega)d^3k$. The volume element is conveniently rescaled as 
\beq d\mu(\bk)=\frac{d^3k}{(2\pi)^32\omega}.\eeq
For example, $\exp(\pm ik\cdot x)$ is a solution provided $k^2+m^2=0$. Now from (\ref{solinnerp}), 
\begin{align}
&\vev{e^{ik\cdot x}|e^{ik'\cdot x}}_{\rm KG}=(2\pi)^32\omega\delta^3(\bk-\bk').\label{expinner1}\\
&\vev{e^{-ik\cdot x}|e^{-ik'\cdot x}}_{\rm KG}=-(2\pi)^32\omega\delta^3(\bk-\bk').\label{expinner2}\\
&\vev{e^{ik\cdot x}|e^{-ik'\cdot x}}_{\rm KG}=0.\label{expinner3}
\end{align}
Although plane wave solutions do not have finite norm wrt (\ref{solinnerp}), we can construct solutions of finite KG-norm from them: For each element $\psi(\bk)\in{\bf L}_2(\bf M_+)$ 
\beq f_\pm(x)=\int_{M_+}e^{\pm ik\cdot x}\psi(\bk)\,d\mu(\bk)\label{statetosol}\eeq
are solutions of KG equation having finite KG-norm $\vev{f_\pm|g_\pm}_{\rm KG}=\pm\vev{\psi|\phi}$ where $\vev{\psi|\phi}$ is the standard ${\bf L}_2(\bf M_+)$ scalar product and $\vev{f_+|g_-}_{\rm KG}=0$. These results follow from (\ref{expinner1}-\ref{expinner3}) directly. The map is invertible, that is for each solution of positive/negative definite KG norm there exists a unique 1-particle state
\beq \psi(\bk)=2\omega(\bk)\int_\Sigma e^{\mp ik\cdot x}f_\pm(x)\,d^3x.\eeq
$f_\pm(x)$ are called the positive and negative frequency solutions of the KG equation. The same calculations also show that if $f(x)$ is a positive frequency solution then its complex conjugate $\overline{f(x)}$ is a negative frequency solution. So a general real solution of KG-equation is $\phi(x)=\sum\alpha_if_i(x)+\overline{\alpha_if_i(x)}$ where $f_i$ is the positive frequency solution associated with a basis $e_i(\bk)$ of the 1-particle Hilbert space $\bf H$ and $\alpha_i$ are some complex numbers. Since each solution is distribution valued, so is $\phi$.

\vspace{0.1cm}
\nd Remarks: The signs of (\ref{expinner1}) and (\ref{expinner2}) depend on our choice of $\epsilon_{0123}=-1$ and Hodge-star operation but the relative sign of (\ref{expinner1}) and (\ref{expinner2}) do not. A different choice will exchange the positive and negative frequency solutions. A positive/negative frequency solution satisfies the equation $i(\del/\del t)f_\pm(x)=\omega f_\pm(x)$ where $\omega$ is positive/negative definite. Furthermore, the time-Fourier transform of a positive frequency solution gives $\int_{-\infty}^\infty dt\exp(i\omega't)f_+(x)=2\pi\delta(\omega-\omega')f_+(0,\bx)$, which vanishes for $\omega'<0$.\vspace{0.1cm}

With each basis state $e_i(\bk)\in{\bf L}_2(\bf M_+)$ we associate an annihilation operator $a_i$ and a creation operator $a^\dagger_i$. The real scalar field operator is defined as follows: A real classical field is the sum of a positive and negative frequency solutions $\phi=\sum\alpha_if_i(x) +\overline{\alpha_if_i(x)}$. Since the complex number $\alpha_i$ is associated with the label of the state $e_i(\bk)$, it is elevated to the operator $a_i$. Similarly $\overline{\alpha_i}$ is elevated to the operator $a^\dagger_i$. So the hermitian scalar field operator $\phi(x)$ is the sum   $\phi(x)=\sum_if_i(x)a_i+\overline{f_i(x)}a^\dagger_i$. Expanding through (\ref{statetosol}),
\beq \phi(x)=\sum_i\int_{M_+}\Big(e^{ik\cdot x}e_i(\bk)a_i+e^{-ik\cdot x}\overline{e_i(\bk)}a^\dagger_i\Big)\,d\mu(\bk)\eeq
Defining $a(\bk)=\sum_ie_i(\bk)a_i$, we get the usual flat spacetime expansions of the free field. However, $[a(\bk),a^\dagger(\bk')]=\sum_ie_i(\bk)\overline{e_i(\bk')}=(2\pi)^32\omega(\bk)\delta^3(\bk-\bk')$.

\vspace{0.1cm}
\nd Remarks: At this stage it is merely a convention that the annihilation operator is associated with the positive frequency solution and the creation operator is associated with the negative frequency solution. This convention fits well with other conventions adopted in canonical quantization. Also note that the field operator depends on the choice of a basis in the 1-particle Hilbert space ${\bf L}_2(\bf M_+)$.
\vspace{0.1cm}

Can we extend the analysis to arbitrary Lorentzian metrics? The scalar product (\ref{solinnerp}) is already defined for an arbitrary metric. Other generalizations are not always possible:

{\tt (a)} The plane waves $\exp(ik\cdot x)$ are eigenfunctions of the wave operator $\del^2$ with negative definite eigenvalues $k^2$. The eigenfunctions of the generalized wave operator $g^{\mu\nu}\grad_{\!\mu}\grad_{\!\nu}$ are unknown. 

{\tt (b)} If in some coordinate system the field equation can be cast into the form 
\beq \frac{\del^2\!\phi}{\del t^2}+K\phi=0\eeq
where $K$ is a second order elliptic differential operator (in flat space $K=-\del_i\del_i+m^2$) that does not involve the time $t$ and is positive definite and a set of eigenfunctions $f_k(\bx)$, where $k$ is an arbitrary label, exists such that
\beq Kf_k(\bx)=\omega^2(k)f_k(\bx),\quad\omega(k)>0,\eeq
then $f_k(\bx)\exp(\mp i\omega(k)t)$ are the positive and negative frequency solutions of the field equation respectively. In the flat case, the KG norm of these solutions (\ref{expinner1})-(\ref{expinner3}) play an important role in choosing the correct Hilbert space ${\bf L}_2(\bf M_+)$, like the correct measure is $d^3k/\omega(\bk)$.

The use of the scalar product (\ref{solinnerp}) is restricted to free fields only because for interacting fields $d*df$ is not proportional to $f$ but to powers of $f$. This prevents the scalar product from being independent of the choice of hypersurface $\Sigma$. One way out of this impasse is to use this scalar product in tangent spaces of the manifold of solutions. In a tangent space, both $f_1,f_2$ are linear perturbations around a solution $f_0$ and satisfy the field equation $(\grad^2-m^2-V''(f_0))f=0$. So the independence is restored. However, this gives a Fock space at each tangent space only. 

\section{Scattering off Black Hole}

The first systematic study of scattering processes in a gravitational field was carried out by Hawking and Wald. They considered a scalar field $\phi_{\rm in}$ in the far past and a field $\phi_{\rm out}$ in the far future when all interactions are turned-off and solutions that interpolates between these fields.
Suppose the two fields are (we are closely following the notations of Wald)
\beq \phi_{\rm in}(x)=\sum G_i(x)a_i+\overline{G_i(x)}a^\dagger_i,\quad\phi_{\rm out}(x)=\sum H_i(x)b_i+\overline{H_i(x)}b^\dagger_i\eeq
and some scattering operator $S$ relates the two fields $S\phi_{\rm in}S^{-1}=\phi_{\rm out}$. 
This implies 
\beq Sa_iS^{-1}=\sum_j\vev{G_i|H_j}_{\rm KG}b_j+\vev{G_i|\overline{H_j}}_{\rm KG}b^\dagger_j.\label{smatrix}\eeq
Now suppose in the far past $H_i$ decomposes into a positive and a negative frequency parts as follows: $H_i=G'_i+\overline{G''_i}$. So while $G_i$ is uniquely associated with the state $e_i\in{\bf H}_{\rm in}$, we suppose $G'_i$ is associated with the state $A_{ij}e_j$ and $G''_i$ is associated with the state $\overline{B_{ij}}e_j$, where $A_{ij},B_{ij}$ are the Bogoliubov coefficients. 

Since $H_i$ is uniquely associated with a state $\tilde e_i\in{\bf H}_{\rm out}$ in the out orthonormal basis, $\vev{H_i|H_j}_{\rm KG}=\vev{\tilde e_i|\tilde e_j}=\delta_{ij}$. So we get (using $\vev{\overline{G_i}|\overline{G_j}}_{\rm KG}=-\vev{e_j|e_i}$)
\begin{align} \delta_{ij}&=\vev{H_i|H_j}_{\rm KG}=\vev{G'_i|G'_j}_{\rm KG}+\vev{\overline{G''_i}|\overline{G''_j}}_{\rm KG}\nn &=\vev{A_{ir}e_r|A_{js}e_s}-\vev{\overline{B_{js}}e_s|\overline{B_{ir}}e_r}=(\overline AA^T-\overline BB^T)_{ij},
\end{align}
that is, $\overline AA^T-\overline BB^T=I$. In a similar manner, we get
\begin{align} 0&=\vev{H_i|\overline{H_j}}_{\rm KG}=\vev{G'_i|G''_j}_{\rm KG}+\vev{\overline{G''_i}|\overline{G'_j}}_{\rm KG}\nn &=\vev{A_{ir}e_r|\overline{B_{js}}e_s}-\vev{A_{js}e_s|\overline{B_{ir}}e_r}=(\overline AB^\dagger-\overline BA^\dagger)_{ij},
\end{align}
that is, $\overline AB^\dagger=\overline BA^\dagger$.

Similarly, supposing that in the far future $G_i$ decomposes into a positive and a negative frequency parts $G_i=H'_i+\overline{H''_i}$ and while $H_i$ is uniquely associated with the state $\tilde e_i\in{\bf H}_{\rm out}$, $H'_i$ is associated with the state $C_{ij}\tilde e_j$ and $H''_i$ is associated with the state $\overline{D_{ij}}\tilde e_j$ where $C_{ij},D_{ij}$ are the Bogoliubov coefficients, we get relations identical to $A,B$: 
\beq \overline CC^T-\overline DD^T=I,\quad\overline CD^\dagger=\overline DC^\dagger.\eeq
Also calculating $\vev{G_i|H_j}_{\rm KG}$ in two different Hilbert spaces ${\bf H}_{\rm in}$ and ${\bf H}_{\rm out}$ we get an additional relation among the Bogoliubov coefficients
\begin{align} \vev{G_i|H_j}_{\rm KG}&=\vev{G_i|G'_j+\overline{G''_i}}_{\rm KG}=\vev{G_i|G'_j}_{\rm KG}=\vev{e_i|A_{jr}e_r}\\
&=\vev{H'_i+\overline{H''_i}|H_j}_{\rm KG}=\vev{H'_i|H_j}_{\rm KG}=\vev{C_{ir}\tilde e_r|\tilde e_j},
\end{align}
which gives $A^T=\overline C$. Similarly, computing $\vev{G_i|\overline{H_j}}_{\rm KG}$ in two difference ways
\begin{align} \vev{G_i|\overline{H_j}}_{\rm KG}&=\vev{G_i|\overline{G'_i}+G''_j}_{\rm KG}=\vev{G_i|G''_j}_{\rm KG}=\vev{e_i|\overline{B_{jr}}e_r}\\
&=\vev{H'_i+\overline{H''_i}|\overline{H_j}}_{\rm KG}=\vev{\overline{H''_i}|\overline{H_j}}_{\rm KG}=-\vev{\tilde e_j|\overline{D_{ir}}\tilde e_r},
\end{align}
we get $B^\dagger=-\overline D$. Taking complex conjugations, the independent relations among all the Bogoliubov coefficient can be re-written as
\begin{align} &AA^\dagger-BB^\dagger=I,\quad AB^T=BA^T,\quad A^\dagger=C,\\
&CC^\dagger-DD^\dagger=I,\quad CD^T=DC^T,\quad B^\dagger=-\overline D.\label{bogocoeff}\end{align}

Using these relations the $S$-matrix equation can be expressed in the matrix form as $SaS^{-1}=A^Tb+B^\dagger b^\dagger$. So if we consider a vacuum state $\Psi_0=(\psi_0,0,0,...)\in{\bf H}_{\rm in}$ then its image state $S\Psi_0\in{\bf H}_{\rm out}$ must satisfy the constraint
\beq SaS^{-1}\!S\Psi_0=Sa\Psi_0=0=(A^Tb+B^\dagger b^\dagger)\Psi,\eeq
which in terms of $C,D$ takes the form $\overline Cb\Psi=\overline Db^\dagger\Psi$. On an arbitrary state (\ref{fockstate}), it gives 
\begin{align} &\overline C_{ij}\Big(\tilde\psi_j,\;\tilde\psi_{jk}\tilde e_k,\;\frac 1{2!}\tilde\psi_{jkl}\tilde e_k\otimes\tilde e_l,...\Big)\nn &=\overline D_{ij}\Big(0,\;\tilde\psi_0\tilde e_j,\;\tilde\psi_k\tilde e_{(j}\otimes\tilde e_{k)},\;\frac 1{2!}\tilde\psi_{kl}\tilde e_{(j}\otimes\tilde e_k\otimes\tilde e_{l)},...\Big).\end{align}
Since $C$ is one-to-one, its inverse exists. Hence this constraint implies $\tilde\psi_i=\tilde\psi_{ijk}=\cdots=0$, that is $\Psi$ may contain only even particle states. This means $\Psi$ is populated with particles created in pairs. 

$S$-matrix gives the following maps
\begin{align} &Sa_iS^{-1}=\vev{G_i|H_s}_{\rm KG}b_s+\vev{G_i|\overline{H_s}}_{\rm KG}b^\dagger_s= A_{si}b_s+\overline{B_{si}}b^\dagger_s,\\
&Sa_i^\dagger S^{-1}=-\vev{\overline{G_i}|H_r}_{\rm KG}b_r-\vev{\overline{G_i}|\overline{H_r}}_{\rm KG}b^\dagger_r=B_{ri}b_r+\overline{A_{ri}}b^\dagger_r.\label{atobbogocoeff}\end{align}
The relations (\ref{bogocoeff}) imply $[Sa_iS^{-1},Sa_j^\dagger S^{-1}]=\delta_{ij}$, and so on.

\nd{\tiny Actually, $[Sa_iS^{-1},Sa_j^\dagger S^{-1}]=(A^T\overline A-B^\dagger\!B)_{ij}$. Putting $A^\dagger=C$ and $B^\dagger=-\overline D$ in the relation $CC^\dagger-DD^\dagger=I$, we get $A^\dagger\!A-B^T\overline B=I$. Taking complex conjugate, we get $A^T\overline A-B^\dagger\!B=I$.}\vspace{0.1cm}

The relations (\ref{atobbogocoeff}) can be inverted
\begin{align} &S^{-1}b_iS=\vev{H_i|G_s}_{\rm KG}a_s+\vev{H_i|\overline{G_s}}_{\rm KG}a^\dagger_s= \overline{A_{is}}a_s-\overline{B_{is}}a^\dagger_s,\\
&S^{-1}b_i^\dagger S=-\vev{\overline{H_i}|G_r}_{\rm KG}a_r-\vev{\overline{H_i}|\overline{G_r}}_{\rm KG}a^\dagger_r=-B_{ir}a_r+A_{ir}a^\dagger_r.\label{btoabogocoeff}\end{align}
These relations imply the image state $S\Psi_0$ measures a total number of particles
\beq \vev{S\Psi_0|b^\dagger_ib_iS\Psi_0}=\vev{\Psi_0|S^{-1}b_i^\dagger SS^{-1}b_iS\Psi_0}={\rm Tr}(BB^\dagger)\eeq
where in the second step we have used $S^\dagger=S^{-1}$, that is $S$-matrix is unitary. The total number of particles is finite iff $B$ is a trace-class operator. 

Suppose a massless scalar test field $\phi$ interacts with gravity when some matter collapses to form an event horizon such that in the far past and future the spacetime is flat. At future null infinity $\mfs I^{\!+}$ a positive frequency solution is $H_\omega\sim\exp(-i\omega u)/r$. We would like to extrapolate this solution to past null infinity $\mfs I^{\!-}$ to see whether we get a $\overline{G''}$. If the angular frequency $\omega\gg 1/r_s$ where $r_s$ is the Schwarzschild radius of the collapsing matter (or the wavelength $\ll r_s$) then the solution may take a null ray back all the way to $\mfs I^{\!-}$. This is possible under two conditions:

{\tt (a)} The null ray does not hit the collapsing matter and gets reflected and/or absorbed by it. So it can take the proposed path only at late times when most of the matter had already crossed the horizon and the ray is not affected by the collapsing matter.

{\tt (b)} The null ray stays outside the even horizon. Since the Kruskal null coordinates are finite close to the event horizon, we should re-express the solution in the Kruskal null coordinate $U=-\exp(-\kappa u)$ where $\kappa$ is the surface gravity of the horizon. We are also assuming a spherically symmetric collapse. 

So $H_\omega\sim\frac 1r(-U)^{i\omega/\kappa}$. In the ray approximation $\omega/\kappa\gg 1$. If the null ray stays outside the event horizon it passes through the origin and traces back to past null infinity where $|U|$ becomes equal to $|v|$. Assuming the last ray from $\mfs I^{\!-}$ reaching $\mfs I^{\!+}$ along the event horizon is emitted at $v=0$, the positive frequency solution of $\mfs I^{\!+}$ extrapolated to $\mfs I^{\!-}$ is $\frac 1r(-v)^{i\omega/\kappa}$. On $\mfs I^{\!-}$ the positive/negative frequency solutions are $\exp(\mp i\omega v)$ respectively. The positive/negative frequency parts of $\frac 1r(-v)^{i\omega/\kappa}$ give $A_{\omega\omega'}$ and $B_{\omega\omega'}$ as follows: 

The KG norm of the positive/negative frequency solutions are
\begin{align}
&\vev{\frac 1re^{-i\omega u}|\frac 1re^{-i\omega'u}}_{\rm KG}=-\vev{\frac 1re^{i\omega u}|\frac 1re^{i\omega'u}}_{\rm KG}=(4\pi)^2\omega\delta(\omega-\omega'),\\
&\vev{\frac 1re^{-i\omega v}|\frac 1re^{-i\omega'v}}_{\rm KG}=-\vev{\frac 1re^{i\omega v}|\frac 1re^{i\omega'v}}_{\rm KG}=(4\pi)^2\omega\delta(\omega-\omega'),\\
&\vev{\frac 1re^{-i\omega u}|\frac 1re^{i\omega'u}}_{\rm KG}=\vev{\frac 1re^{-i\omega v}|\frac 1re^{i\omega'v}}_{\rm KG}=0.
\end{align}
The map between the Hilbert space and positive frequency solutions is
\beq H_k(x)=\int_0^\infty\frac{\exp(-i\omega u)}r\frac{L_k(\omega l)}{k!}\sqrt{\omega l}e^{-\omega l/2}\frac{d\omega}{4\pi\omega}\eeq
where $\exp(-x/2)L_k(x)/k!$, $k=0,1,2,...$, are the orthnormalized Laguerre polynomials in ${\bf L}_2(0,\infty)$ and $l$ is some arbitrary length scale. By construction, $H_k$ are orthonormal in the KG-norm.

The same solutions extrapolated to $\mfs I^{\!-}$ are obtained by replacing $\exp(-i\omega u)$ in $H_k(x)$ by $(-v)^{i\omega/\kappa}$. The positive frequency solutions at $\mfs I^{\!-}$, denoted by $G_k$ are obtained by replacing $\exp(-i\omega u)$ in $H_k$ by $\exp(-i\omega v)$. $H_k$ is decomposed into $G_k'$ and $\overline{G_k''}$. So the Bogoliubov coefficients are
\begin{align}
&A_{ks}=\vev{G_s|H_k}_{\rm KG}=\vev{e_s|Ae_k}=\int_0^\infty d\omega d\omega'\vev{e_s|\omega'}A_{\omega\omega'}\vev{\omega|e_k}\\
&B_{ks}=-\vev{\overline{G_s}|H_k}_{\rm KG}=\vev{e_s|Be_k}=\int_0^\infty d\omega d\omega'\vev{e_s|\omega'}B_{\omega\omega'}\vev{\omega|e_k}
\end{align}
where $\vev{e_s|\omega}=(L_s(\omega l)/s!)\sqrt l\exp(-\omega l/2)$. Calculating the KG-norms, we get
\beq 
A_{\omega\omega'}=\frac 1{2\pi}\sqrt{\frac{\omega'}{\omega}}\frac{\Gamma(1+i\omega/\kappa)}{(i\omega')^{1+i\omega/\kappa}},\quad
B_{\omega\omega'}=\frac 1{2\pi}\sqrt{\frac{\omega'}{\omega}}\frac{\Gamma(1+i\omega/\kappa)}{(-i\omega')^{1+i\omega/\kappa}},\label{bogocoeffb}
\eeq
%
%
It shows $A_{\omega\omega'}=-B_{\omega\omega'}\exp(\pi\omega/\kappa)$. Finally the number of particles with frequency $\omega$ is obtained from the relation $(AA^\dagger-BB^\dagger)_{\omega\omega}=1$
\beq N_\omega=\frac 1{\exp(2\pi\omega/\kappa)-1}\eeq
which comparing with the Bose-Einstein distribution gives a black body temperature $T=\kappa/2\pi$ called the Hawking temperature.

\section{Local calculations}

It is worth noting that an equivalent calculation may be carried out locally in some neighborhood of the horizon. A precise definition of this neighborhood as an asymptotically flat region of spacetime may be given in a similar way some neighborhoods of the future and past null infinities are identified as asymptotically flat regions. In a spherically symmetric collapse the metric is static and regular at the horizon at sufficiently late times in appropriate coordinates,
\beq ds^2=-F^2(U,V)\,dUdV+r^2d\Omega\eeq
where the metric coefficient $F^2(U,V)$ has a regular limit $\alpha^2$ as $U,V\to 0$. The effective line-element is $ds^2=-\alpha^2dUdV+r_s^2d\Omega$. In these coordinates, the plane S-wave solutions are $\exp(-i\omega U/\kappa)$ and/or $\exp(-i\omega V/\kappa)$, which are positive frequency eigenmodes wrt the timelike vector field $i\del/\del T$ where $T=\alpha(U+V)/2$. However, the eigenmodes wrt the timelike Killing vector field $i(-U\del_U+V\del_V)$ is $U^{i\omega/\kappa}$ and/or $V^{-i\omega/\kappa}$. The KG-norms on $T={\rm constant}$ slices are
\begin{align}
&\vev{\frac 1{r_s}e^{-i\omega U/\kappa}|\frac 1{r_s}e^{-i\omega'U/\kappa}}_{\rm KG}=-\vev{\frac 1{r_s}e^{i\omega U/\kappa}|\frac 1{r_s}e^{i\omega'U/\kappa}}_{\rm KG}=(4\pi)^2\omega\delta(\omega-\omega'),\\
&\vev{\frac 1{r_s}e^{-i\omega V/\kappa}|\frac 1{r_s}e^{-i\omega'V/\kappa}}_{\rm KG}=-\vev{\frac 1{r_s}e^{i\omega V/\kappa}|\frac 1{r_s}e^{i\omega'V/\kappa}}_{\rm KG}=(4\pi)^2\omega\delta(\omega-\omega'),\\
&\vev{\frac 1{r_s}e^{-i\omega U/\kappa}|\frac 1{r_s}e^{i\omega'U/\kappa}}_{\rm KG}=\vev{\frac 1{r_s}e^{-i\omega V/\kappa}|\frac 1{r_s}e^{i\omega'V/\kappa}}_{\rm KG}=0.
\end{align}
The map between the Hilbert space and positive frequency solutions is
\beq H_n(x)=\int_0^\infty\frac{\exp(-i\omega V/\kappa)}{r_s}\frac{L_n(\omega/\kappa)}{n!}e^{-\omega/2\kappa}\frac{d\omega}{4\pi\sqrt{\omega\kappa}}\eeq
where $H_n$ are orthonormal in the KG-norm. Note that the positive and negative frequency eigenfunctions of the timelike Killing vector, respectively given by $(-U)^{i\omega/\kappa}$ and $V^{-i\omega/\kappa}$, have the same KG-norms on $T={\rm constant}$ slices, namely
\begin{align}
&\vev{\frac 1{r_s}(-U)^{i\omega/\kappa}|\frac 1{r_s}(-U)^{i\omega'/\kappa}}_{\rm KG}=-\vev{\frac 1{r_s}(-U)^{-i\omega/\kappa}|\frac 1{r_s}(-U)^{-i\omega'/\kappa}}_{\rm KG}=(4\pi)^2\omega\delta(\omega-\omega'),\nn
&\vev{\frac 1{r_s}V^{-i\omega/\kappa}|\frac 1{r_s}V^{-i\omega'/\kappa}}_{\rm KG}=-\vev{\frac 1{r_s}V^{i\omega/\kappa}|\frac 1{r_s}V^{i\omega'/\kappa}}_{\rm KG}=(4\pi)^2\omega\delta(\omega-\omega'),\nn
&\vev{\frac 1{r_s}(-U)^{i\omega/\kappa}|\frac 1{r_s}(-U)^{-i\omega'/\kappa}}_{\rm KG}=\vev{\frac 1{r_s}V^{-i\omega/\kappa}|\frac 1{r_s}V^{i\omega'/\kappa}}_{\rm KG}=0.
\end{align}

The same solutions of $\mfs I^{\!+}$ extrapolated to the near-horizon region are obtained by replacing $(-U)^{i\omega/\kappa}$ by $V^{i\omega/\kappa}$. The positive frequency solutions $H_k$ associated with the local time coordinate is then decomposed into $G_k'$ and $\overline{G_k''}$, the positive and negative frequency solutions associated with the Killing time. The Bogoliubov coefficients are
\begin{align}
&A_{ks}=\vev{G_s|H_k}_{\rm KG}=\vev{e_s|Ae_k}=\int_0^\infty d\omega d\omega'\vev{e_s|\omega'}A_{\omega\omega'}\vev{\omega|e_k}\\
&B_{ks}=-\vev{\overline{G_s}|H_k}_{\rm KG}=\vev{e_s|Be_k}=\int_0^\infty d\omega d\omega'\vev{e_s|\omega'}B_{\omega\omega'}\vev{\omega|e_k}
\end{align}
where $\vev{e_s|\omega}=(L_s(\omega l)/s!)\sqrt l\exp(-\omega l/2)$ for $l=1/\kappa$. Calculating the KG-norms, we get
\beq 
A_{\omega\omega'}=\frac 1{2\pi}\sqrt{\frac{\omega'}{\omega}}\frac{\Gamma(1+i\omega/\kappa)}{(i\omega'/\kappa)^{1+i\omega/\kappa}},\quad
B_{\omega\omega'}=\frac 1{2\pi}\sqrt{\frac{\omega'}{\omega}}\frac{\Gamma(1+i\omega/\kappa)}{(-i\omega'/\kappa)^{1+i\omega/\kappa}},
\eeq
which conforms with the earlier coefficients (\ref{bogocoeffb}).

\section{Conclusion}

Although the calculation of Bogoliubov coefficients are similar, there are important differences between the local and global calculations. First, the local calculations are carried out in a neighborhood of the horizon. Since the eigenmodes of a timelike Killing vector is used in an essential way, the horizon is necessarily of a Killing type. This feature may not play an important role in the global calculations because at flat asymptotic infinity one has a preferred timelike vector. Second, the space-time in this neighborhood is approximately flat is also used in an essential way. This is also a feature of only a special class of black hole solutions that are of non-extremal type. Finally, the calculations can hardly be interpreted in the light of scattering theory because the gravitational interactions do not fall-off as expected in a scattering theory like in asymptotic infinities. On the contrary, the local calculation depends entirely on the existence of two sets of creation-annihilation operators, one associated with a local regular time $T$, and another associated with the Killing time. The choice of the local time $T$ is somewhat arbitrary but the $A,B$ coefficients do not depend on this choice. This freedom is built-in in the KG scalar product and closely related to the independence of KG scalar product on the choice of spacelike hypersurfaces $\Sigma$. It explicitly demonstrates, how without using the asymptotic structures of space-time, one can use local characteristics of Killing horizon to extract the results of Hawking and Wald. To the best of my knowledge, such a calculation was suggested in \cite{Wald2} along the line of Hartle-Hawking vacuum, but was not explicitly carried out before.

An important question remains: Can we dispose off the Killing vector from local calculations? If it is possible then a larger family of black hole solutions can be incorporated in the present scheme. 

\section{Acknowledgments}

I am fortunate to present these calculations in some talks given at the Central University of Himachal Pradesh and S N Bose National Center of Basic Sciences, Kolkata. I gratefully acknowledge discussions with Dr. Ayan Chatterjee, Dr. Avirup Ghosh and Pritam Nanda.

\end{document}